\documentclass[a4paper,fleqn]{cas-sc}

\usepackage[numbers]{natbib}
\usepackage{setspace}
\usepackage{comment}

\def\tsc#1{\csdef{#1}{\textsc{\lowercase{#1}}\xspace}}
\tsc{WGM}
\tsc{QE}
\tsc{EP}
\tsc{PMS}
\tsc{BEC}
\tsc{DE}

\begin{document}
\let\WriteBookmarks\relax
\def\floatpagepagefraction{1}
\def\textpagefraction{.001}
\shorttitle{L{\'e}vy noise-induced coherence resonance}
\shortauthors{I. Korneev, A. Zakharova and V.Semenov}

\title [mode = title]{L{\'e}vy noise-induced coherence resonance: numerical study versus experiment}                      

\author[1]{Ivan Korneev}[orcid=0000-0003-1716-5170]
\address[1]{Institute of Physics, Saratov State University, 83 Astrakhanskaya str., 410012 Saratov, Russia}

\author[2]{Anna Zakharova}[orcid=0000-0002-1499-6043]
\address[2]{Bernstein Center for Computational Neuroscience, Humboldt-Universit\"{a}t zu Berlin, Philippstra{\ss}e 13, 10115 Berlin, Germany
}

\author[1]{Vladimir V. Semenov}[orcid=0000-0002-4534-8065]
\corref{cor1}


\cortext[cor1]{Corresponding author}

\begin{abstract}
Using the FitzHugh-Nagumo system in the excitable regime, we investigate the influence of the L{\'e}vy noise properties on the effect of coherence resonance. In particular, we demonstrate that the L{\'e}vy noise can be a constructive or destructive factor providing for enhancement or suppression of noise-induced coherence. We show that the positive or negative role of the L{\'e}vy noise impact is dictated by the noise stability index and skewness parameter. The correlation time and the deviation of interspike intervals used in this analysis are shown to be maximized or minimized for an appropriate choice of the noise parameters. Numerical simulations are combined with experiments on an electronic circuit showing an excellent qualitative correspondence and proving thereby the robustness of the observed phenomena.
\end{abstract}



\maketitle

\doublespacing

\section{\label{sec:intro}Introduction}

First coined by Pikovsky and Kurths \cite{pikovsky1997}, coherence resonance has become an interdisciplinary phenomenon observed in a wide range of excitable \cite{pikovsky1997,lindner1999,lindner2004,deville2005,muratov2005,semenov2017,semenov2018} and non-excitable \cite{gang1993,ushakov2005,zakharova2010,zakharova2013,geffert2014,semenov2015} systems. The essence of coherence resonance is increasing the noise-induced oscillation regularity for an optimum value of the noise intensity. Such effects are a frequent occurrence in neurodynamics \cite{pikovsky1997,lee1998,lindner2004,tateno2004} as well as in microwave \cite{dmitriev2011} and semiconductor \cite{hizanidis2006,huang2014,shao2018} electronics, optics \cite{giacomelli2000,avila2004,otto2014,arteaga2007,arecchi2009}, quantum physics \cite{kato2021}, thermoacoustics \cite{kabiraj2015}, plasma physics \cite{shaw2015}, hydrodynamics \cite{zhu2019}, climatology \cite{bosio2023} and chemistry \cite{miyakawa2002,beato2008,simakov2013}.

The issues addressing coherence resonance involve a broad spectrum of dynamical systems and practical applications from laser pulse reshaping and clock recovery \cite{dubbeldam1999} to noise-enhanced performance of neurointerfaces (many of them are discussed in review \cite{pisarchik2023}). For this reason, the problem of coherence resonance control is relevant in the context of both fundamental and applied science. There are known several methods for controlling the characteristics of noise-induced oscillations in the regime of coherence resonance. For this purpose, one can introduce time-delayed feedback. This approach is useful to control coherence resonance in systems with type-I \cite{aust2010} and type-II \cite{janson2004,brandstetter2010} excitability as well as in non-excitable systems \cite{geffert2014,semenov2015}. If the coherence resonance occurs in networks of coupled oscillators, one can vary the coupling strength to control this effect. As an example, such strategy is applied for coherence resonance control in multilayer network with multiplexing \cite{semenova2018,masoliver2021}. In the current paper, we complement a manifold of coherence resonance control schemes and demonstrate that the coherence resonance in excitable systems can be efficiently controlled by varying the L{\'e}vy noise parameters. 

L{\'e}vy noise is a class of stable non-Gaussian noise that exhibits long heavy tails of its distribution associated with large, potentially infinite, jumps. L{\'e}vy noise is often used to model stochastic systems where the noise-induced dynamics is characterised by abrupt changes. One can observe such effects in lasers \cite{rocha2020}, cardiac rhythms \cite{peng1993}, molecular motors \cite{lisowski2015}, quantum dots \cite{novikov2005}, financial \cite{mantegna1999,barndorff2001} and social \cite{perc2007} systems. Moreover, stochastic processes with a L{\'e}vy distribution can model the dynamics of real biological neurons more accurately as compared to Gaussian noise \cite{nurzaman2011,wu2017}. Motivated by the significance of L{\'e}vy processes in neural systems, in this paper we are focused on coherence resonance induced by L{\'e}vy noise in the FitzHugh-Nagumo oscillator being a classical model of a spiking neuron. 

In the context of noise-induced resonant phenomena, L{\'e}vy noise is known to provide for controlling characteristics of noise-induced oscillations in the regime of stochastic resonance \cite{dybiec2006,dybiec2009,yonkeu2020}, self-induced stochastic resonance \cite{yamakou2022}, coherence resonance in non-excitable systems \cite{yonkeu2020}. In the present paper, we extend this list by one more effect, coherence resonance in an excitable system where additive L{\'e}vy noise induces more or less pronounced coherence resonance when varying the noise characteristics being responsible for extreme jump generation. We combine methods of numerical simulation and experimental research by using an electronic model of the FitzHugh-Nagumo oscillator subject to L{\'e}vy noise generated by a personal-computer-based system.

\section{\label{sec:model}Model and methods}
The FitzHugh-Nagumo oscillator \cite{fitzhugh1961,nagumo1962} is a paradigmatic model for the type-II excitability. In the present paper, we consider the FitzHugh-Nagumo model in the simplest form:
\begin{equation}
\label{eq:system} 
\begin{array}{l} 
\varepsilon\dfrac{dx}{dt} = x-x^3/3-y, \quad \dfrac{dy}{dt}=x+a+\xi(t), 
\end{array}
\end{equation} 
where $x=x(t)$ and $y=y(t)$ are dynamical variables. A parameter $\varepsilon\ll 1$ is responsible for the time scale separation of fast activator, $x$, and slow inhibitor, $y$, variables, $a$ is the threshold parameter which determines the system dynamics: the system exhibits the excitable regime at $|a|>1$ and the oscillatory one for $|a|<1$. In this paper, we consider the FitzHugh-Nagumo system in the excitable regime ($a=1.05$ and $\varepsilon=0.05$) for varying parameters of additive L{\'e}vy noise $\xi(t)$, which is defined as the formal derivative of the L{\'e}vy stable motion. L{\'e}vy noise is characterized by four parameters: stability index $\alpha \in (0:2]$, skewness (asymmetry) parameter $\beta\in [-1:1]$, mean value $\mu=0$ ($\mu$ is set to be zero for the strictly stable distributions \cite{janicki1994}) and scale parameter $\sigma$. Parameter $D=\sigma^{\alpha}$ is introduced as the noise intensity. If $\xi(t)$ obeys to L{\'e}vy distribution $L_{\alpha,\beta}(\xi,\sigma,\mu)$, its characteristic function takes the form \cite{janicki1994,dybiec2006,dybiec2007}:
\begin{equation}
\label{eq:characteristic_function} 
\phi(k)=\int\limits_{-\infty}^{+\infty}\exp(ikx)L_{\alpha,\beta}(\xi,\sigma,\mu)dx=
\left\lbrace
\begin{array}{l} 
\exp\left[-\sigma^{\alpha}|k|^{\alpha}\left(1-i\beta sgn(k)\tan\dfrac{\pi\alpha}{2} \right) \right],\quad \text{for } \alpha\neq 1,\\ 
\exp\left[-\sigma |k|\left(1+i\beta\dfrac{2}{\pi} sgn(k)\ln|k| \right) \right],\quad \text{for } \alpha=1.
\end{array}
\right.
\end{equation} 
To generate random sequence $\xi$ corresponding to characteristic function (\ref{eq:characteristic_function}), the Janicki-Weron algorithm is used \cite{janicki1994,weron1995}: 
\begin{equation}
\label{eq:noise_generation} 
\begin{array}{l} 
\xi=\sigma S_{\alpha,\beta}\times \dfrac{\sin(\alpha(V+B_{\alpha,\beta}))}{(\cos(V))^{1/\alpha}}\times \left( \dfrac{\cos(V-\alpha(V+B_{\alpha,\beta}))}{W}\right)^{\dfrac{1-\alpha}{\alpha}}, \quad \text{for }\alpha\neq 1,\\
\xi=\dfrac{2\sigma}{\pi}\left[ \left( \dfrac{\pi}{2}+\beta V \right) \tan(V)-\beta \ln \left(\dfrac{\dfrac{\pi}{2}W\cos(V)}{\dfrac{\pi}{2}+\beta V}\right)\right], \quad \text{for }\alpha=1,
\end{array}
\end{equation} 
where $B_{\alpha,\beta}=\left( \text{arctan} \left( \beta \tan \left( \dfrac{\pi\alpha}{2}\right)\right) \right)/\alpha$, $S_{\alpha,\beta}=\left( 1+\beta^2 \tan ^2\left( \dfrac{\pi\alpha}{2}\right)\right)^{1/2\alpha}$, $V$ is a random variable uniformly distributed on $\left(-\dfrac{\pi}{2}:\dfrac{\pi}{2}\right)$, $W$ is an exponential random variable with mean 1 (variables $W$ and $V$ are independent). In case $\alpha=2$, the distribution of the probability density function (PDF) takes the Gaussian form with zero mean value and the variance being equal to $2\sigma^2$. If $\alpha<2$, the distribution is non-Gaussian and the variance is infinite.

Numerical simulations of model equations  (\ref{eq:system}) are carried out by the integration by the Heun method \cite{mannella2002} with the time step $\Delta t=10^{-3}$ or smaller. It is important to note that numerical modelling of equations including $\alpha$-stable stochastic process with finite time step implies the normalization of the noise term by $\Delta t^{1/\alpha}$ \cite{xu2016,pavlyukevich2010}.

It is well-known that methods of numerical simulations of stochastic differential equations can accumulate errors and give rise to wrong results in the presence of high-level noise or close to bifurcation transitions. Dynamical systems subject to random perturbations including extreme jumps are also difficult to model. To avoid misleading conclusions and confirm results of numerical simulations, numerical experiment is combined with the physical one. The physical experiment was carried out with using an experimental prototype being an electronic model of system (\ref{eq:system}) implemented by principles of analog modelling \cite{luchinsky1998}. Detailed circuit diagram of the experimental setup is shown in Fig. \ref{fig1}. It contains two integrators, A1 and A2, whose output voltages are taken as the dynamical variables, $x$ and $y$, respectively. The experimental facility includes a source of L{\'e}vy noise, which was implemented using a personal computer complemented by an electronic board (National Instruments NI-PCI 6251). To create the L{\'e}vy noise source, an algorithm of random number generation (\ref{eq:noise_generation}) is realized on the Labview platform. The Labview program includes a block being responsible for generating random numbers $\xi$ with high frequency as an output analog signal (voltage) of the NI-board. Signal $\xi(t)$ in Fig. \ref{fig1} is assumed to be white noise, since the  sample clock rate of random number generator (up to $10^6$ samples per second) is much higher than the main oscillation frequency of the experimental setup. Operation of the experimental setup is described by the following equations:
\begin{figure}[t]
\centering
\includegraphics[width=0.5\textwidth]{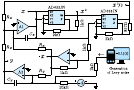}
\caption{Circuit diagram of the experimental setup (see Eqs. (\ref{eq:exp_setup})). Operational amplifiers are TL072CP. Analog integrator elements are $C_x=10$ nF, $R_x=1$ k$\Omega$, $C_y=20$ nF, $R_y=10$ k$\Omega$, parameter $a$ is fixed, $a=1.05$.}
\label{fig1}
\end{figure}  
\begin{equation}
\label{eq:exp_setup} 
\begin{array}{l} 
R_xC_x\dfrac{dx}{dt} = x-x^3/3-y, \quad R_yC_y\dfrac{dy}{dt}=x+a+\xi(t), 
\end{array}
\end{equation} 
where $C_x=10$ nF, $R_x=1$ k$\Omega$, $C_y=20$ nF, $R_y=10$ k$\Omega$ are the resistances and capacitances at the integrators A1 and A2, correspondingly. The input signals of integrator A2 are: signal $-y$ (arrives at the resistor $R_y$), signal $-a$ (goes through the resistance $R_y=10$ k$\Omega$), and signal $-0.1\xi(t)$ (applied to the resistor $0.1R_y=1$ k$\Omega$ to multiply the noise term by $10$ in Eqs. (\ref{eq:exp_setup})). This provides for more accurate tuning parameter $a$ and minimizing the amplitude limitation of the signals generated by the NI-board. In particular, the amplitude of L{\'e}vy noise jumps is much higher than the operating range of the NI-board's output voltage $[-10$ V$:10$ V]. Connection through the resistance $0.1R_y$ allows to realize signal $\xi(t)$, whose amplitude exceeds values in the range  $[-100:100]$. Equations of the experimental setup (\ref{eq:exp_setup}) can be transformed into dimensionless system (\ref{eq:system}) with $\varepsilon=(R_xC_x)/(R_yC_y)=0.05$ by using substitution $t=t/\tau_0$ ($\tau_0=R_yC_y=0.2$ ms is the circuit's time constant) and new dynamical variables $x/V_0$ and $y/V_0$, where $V_0$ is the unity voltage, $V_0=1$ V. 

To reveal the intrinsic properties of coherence resonance induced by the L{\'e}vy noise, we explore the evolution of time realizations and phase portraits caused by the noise intensity growth. In addition, we analyze the dependencies of the correlation time, $t_{\text{cor}}$, and the normalized standard deviation of interspike intervals, $R_T$, introduced in the form:
\begin{equation}
\label{eq:t_cor_and_RT} 
t_{\text{cor}}=\dfrac{1}{\Psi(0)}\int\limits_{0}^{\infty} \left| \Psi(s) \right|ds, \quad R_{T}=\dfrac{\sqrt{\left< T_{ISI}^2\right>-\left< T_{ISI}\right>^2}}{\left< T_{ISI}\right>},
\end{equation} 
where $\Psi(s)$ and $\Psi(0)$ are the autocorrelation function and the variance of the time realization $x(t)$, $T_{ISI}$ is a sequence of time intervals between neighbour spikes. In the following, the evolution of the dynamics caused by increasing noise level is described by using dependencies of $t_{\text{cor}}$ and $R_T$ on the noise intensity, $D$. Mathematical model (\ref{eq:system}) and experimental setup equations (\ref{eq:exp_setup}) have different time scales, which affects the algorithm for the experimental noise intensity calculation. To illustrate the dependencies obtained by means of numerical modelling ($t_{\text{cor}}(D)$ and $R_T(D)$) and physical experiments ($(R_yC_y)^{-1} t_{\text{cor}}(D_{\text{exp}})$ and $R_T(D_{\text{exp}})$) in the same scale, the experimental noise intensity $D_{\text{exp}}$ is introduced as $D_{\text{exp}}=\sigma^{\alpha}\tau_0^{\left(1-\dfrac{1}{\alpha}\right)}$.

\section{\label{sec:tails}The impact of L{\'e}vy jumps}
\begin{figure}[t!]
\centering
\includegraphics[width=1.00\textwidth]{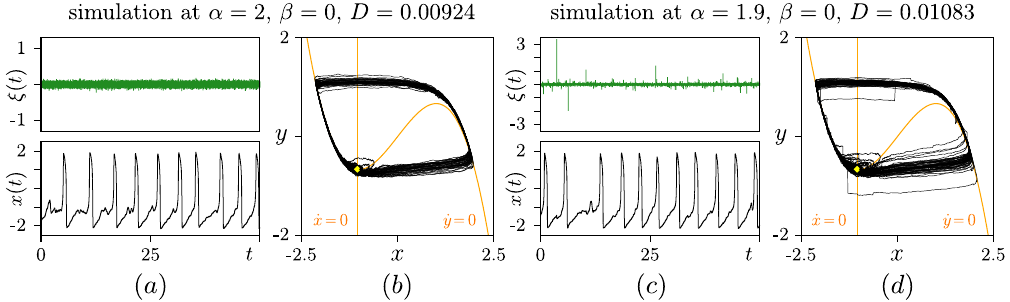}
\caption{Noise-induced oscillations observed in numerical model (\ref{eq:system}) at $\alpha=2$ (Gaussian noise) and $\alpha=1.9$ (non-Gaussian noise). Panels (a) and (c) illustrate oscillations $\xi(t)$ (upper panels) and $x(t)$ (lower panels), whereas panels (b) and (d) depict the phase portraits in the plane ($x$, $y$). Orange curves are nullclines $\dfrac{dx}{dt}=0$ and $\dfrac{dy}{dt}=0$, the yellow point is a stable fixed point. System parameters are: $a=1.05$, $\varepsilon=0.05$, $\beta=0$.}
\label{fig2}
\end{figure}

\begin{figure}[t!]
\centering
\includegraphics[width=1.00\textwidth]{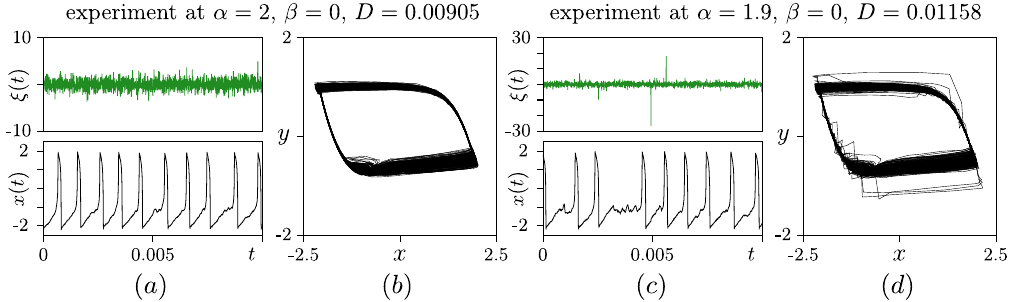}
\caption{Noise-induced oscillations observed in physical experiments (see Eqs. (\ref{eq:exp_setup})) at $\alpha=2$ (Gaussian noise) and $\alpha=1.9$ (non-Gaussian noise). Panels (a) and (c) illustrate oscillations $\xi(t)$ (upper panels) and $x(t)$ (lower panels), whereas panels (b) and (d) depict the phase portraits in the plane ($x$, $y$). System parameters are the same as in Fig. \ref{fig2}.}
\label{fig3}
\end{figure}

In case $\alpha=2$, L{\'e}vy jumps are absent and the PDF of $\xi(t)$ takes the form of Gaussian distribution. Increasing the noise intensity gives rise to more and more frequent noise-induced spikes (depicted in Fig.~\ref{fig2}~(a) and Fig.~\ref{fig3}~(a)). The corresponding trajectories on the phase plane ($x$,$y$) trace oscillations of the phase point near the limit cycle exhibited by model (\ref{eq:system}) and experimental setup (\ref{eq:exp_setup}) at $|a|<1$ (see Fig.~\ref{fig2}~(b) and Fig.~\ref{fig3}~(b)). Then one deals with coherence resonance manifested in the classical well-studied form, in the presence of white Gaussian noise. 

When parameter $\alpha$ decreases, spontaneous L{\'e}vy jumps appear [Fig.~\ref{fig2}~(c) and Fig.~\ref{fig3}~(c)]. As a result, the shape of noise-induced oscillations on the phase plane becomes distorted and the phase trajectories are more 'scattered', especially in the vertical direction [Fig.~\ref{fig2}~(d) and Fig.~\ref{fig3}~(d)]. Intriguingly, this effect has no visible impact on the spikes'form (compare realizations $x(t)$ in Fig.~\ref{fig2}~(a),(c) and Fig.~\ref{fig3}~(a),(c)). If parameter $\beta$ equals to zero, jumps up and down of noise signal $\xi(t)$ are equally probable and its PDF is symmetric. The jumps of signal $\xi(t)$ are high-amplitude short impulses occasionally kicking the phase point out of the phase space area where the noise-induced oscillations are observed in presence of Gaussian noise [Fig.~\ref{fig2}~(d) and Fig.~\ref{fig3}~(d)]. This effect becomes stronger when decreasing parameter $\alpha$. Analysing the phase portraits obtained in numerical simulations [Fig.~\ref{fig2}~(b),(d)] and physical experiment [Fig.~\ref{fig3}~(b),(d)] for identical noise parameters, one can conclude that the noise-induced oscillations become less regular with decrease of $\alpha$. However, this statement requires more detailed examination than studying time realizations and phase portraits.

\begin{figure}[t!]
\centering
\includegraphics[width=1.00\textwidth]{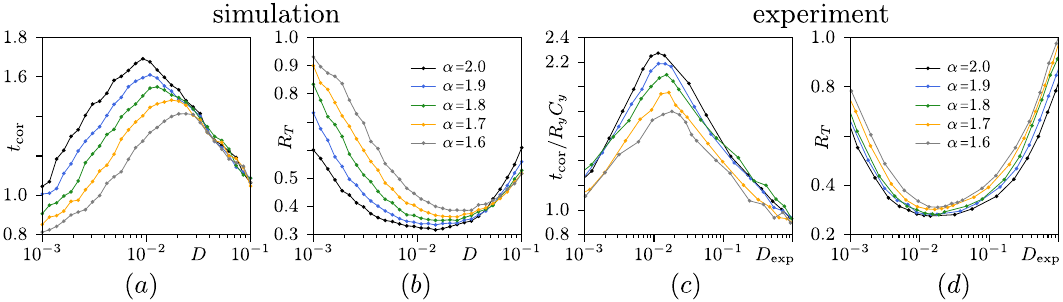}
\caption{Coherence resonance observed in numerical (panels (a) and (b)) and physical (panels (c) and (d)) experiments when decreasing parameter $\alpha$. The effect is illustrated by the dependencies of the correlation time (panels (a) and (c)) and the inter-spike interval deviation (panels (b) and (d)) on the noise intensity. The system parameters are the same as in Fig. \ref{fig2} except of parameter $\alpha$.}
\label{fig4}
\end{figure}  

To prove that symmetric L{\'e}vy jumps decrease the oscillation regularity in the regime of coherence resonance, we take into consideration the evolution of the correlation time, $t_{\text{cor}}$, and the deviation of interspike intervals, $R_T$, caused by changing the noise parameters. Indeed, the described in the previous paragraph effect of L{\'e}vy jumps results in the suppression of coherence resonance reflected in the dependencies $t_{\text{cor}}$ and $R_T$ on the noise intensity [Fig.~\ref{fig4}]. The effect of coherence resonance becomes less and less pronounced with decreasing parameter $\alpha$. It is indicated both in numerical simulations and physical experiments that the decrease of $\alpha$ reduces the highest possible values of $t_{\text{cor}}$, whereas the minimal values of $R_T$ become higher. Observed in physical experiments, these effects are less pronounced, but still visible. This is due to limitation of noise signal amplitude and individual peculiarities of the experimental setup.

\section{\label{sec:skewness}The influence of skewness}

\begin{figure}[b]
\centering
\includegraphics[width=1.00\textwidth]{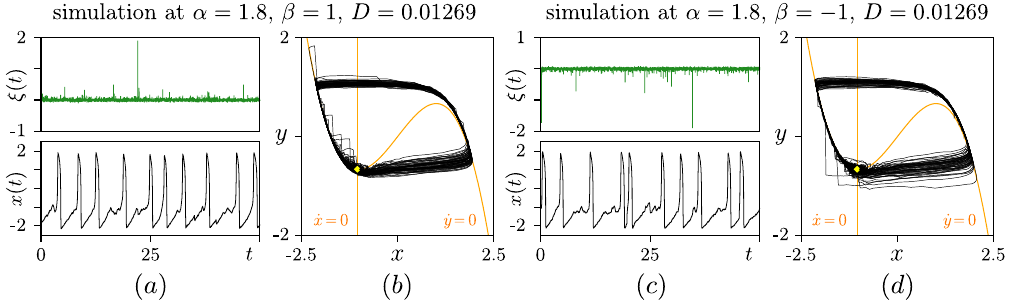}
\caption{Impact of the L{\'e}vy noise skewness on the noise-induced oscillations indicated in numerical model (\ref{eq:system}) at $\beta=1$ and $\beta=-1$. Panels (a) and (c) illustrate oscillations $\xi(t)$ (upper panels) and $x(t)$ (lower panels), whereas panels (b) and (d) depict the phase portraits in the plane ($x$, $y$). The nullclines and stable steady state are coloured as in Fig. \ref{fig2}. System parameters are: $a=1.05$, $\varepsilon=0.05$, $\alpha=1.8$.}
\label{fig5}
\end{figure}  
\begin{figure}[b]
\centering
\includegraphics[width=1.00\textwidth]{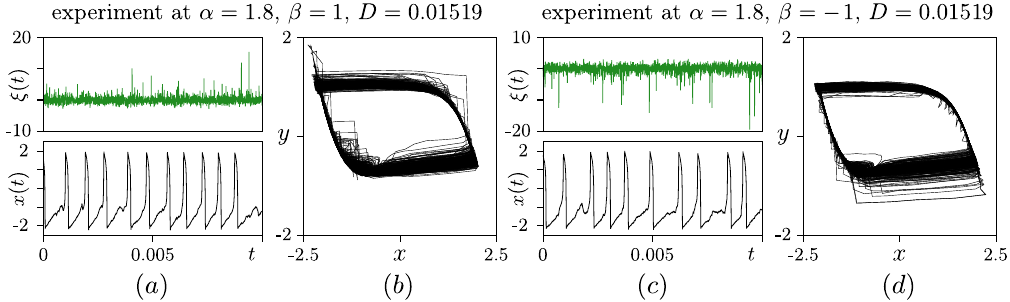}
\caption{Impact of the L{\'e}vy noise skewness on the noise-induced oscillations registered in electronic experiments (see Eqs. (\ref{eq:exp_setup})) at $\beta=1$ and $\beta=-1$. Panels (a) and (c) illustrate oscillations $\xi(t)$ (upper panels) and $x(t)$ (lower panels), whereas panels (b) and (d) depict the phase portraits in the plane ($x$, $y$). System parameters are the same as in Fig. \ref{fig5}.}
\label{fig6}
\end{figure}  

\begin{figure}[t]
\centering
\includegraphics[width=1.00\textwidth]{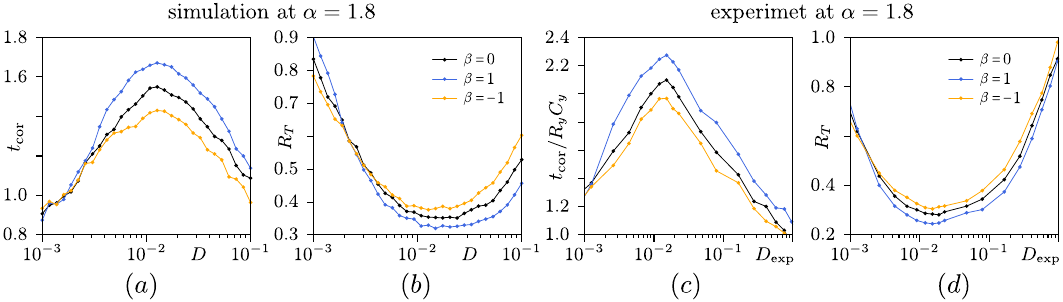}
\caption{Coherence resonance in the presence of the L{\'e}vy noise skewness observed in numerical (panels (a) and (b)) and physical (panels (c) and (d)) experiments. The effect is illustrated by the dependencies of the correlation time (panels (a) and (c)) and the inter-spike interval deviation (panels (b) and (d)) on the noise intensity. The system parameters are the same as in Fig. \ref{fig5} except of $\beta$.}
\label{fig7}
\end{figure}

Next, we focus on the impact of the noise skewness, $\beta\neq0$. The stability index is fixed, $\alpha=1.8$. In such a case, the noise signal $\xi$ is asymmetric and either jumps up or jumps down are more pronounced which also has no principal impact on the shape of the noise-induced spikes (see realizations in Fig.~\ref{fig5}~(a),(c) and Fig.~\ref{fig6}~(a),(c)). However, vertical jumps of the phase point are visible on the phase plane (see the phase portraits in Fig.~\ref{fig5}~(b),(d) and Fig.~\ref{fig6}~(b),(d)). During numerical and experimental researches, we found that varying the L{\'e}vy noise parameters in the presence of skewness provides for the observation of a wider variety of effects in comparison with case $\beta=0$. In particular, depending on the sign of parameter $\beta$, one can achieve suppressed or enhanced coherence resonance. The ability to control the noise-induced coherence is visualised in Fig. \ref{fig7}. It has been established both in numerical and physical experiments that coherence resonance observed at positive values $\beta$ is characterised by increasing the maximal correlation time and decreasing the minimal deviation of interspike intervals achieved at intermediate values of the noise intensity. In contrast, negative values of $\beta$ give rise to the opposite effect: one achieves the lower correlation time and higher deviation of interspike intervals.
\begin{figure}[t]
\centering
\includegraphics[width=0.50\textwidth]{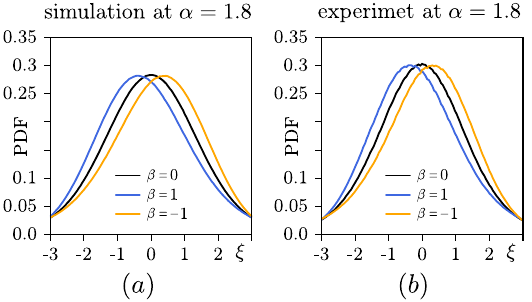}
\caption{Evolution of the PDF characterising the noise signal $\xi(t)$ (see Eqs. \ref{eq:noise_generation}) when varying parameter $\beta$. Other noise parameters: $\alpha=1.8$, $\sigma=1$. }
\label{fig8}
\end{figure}  

The reason for the noise-induced dynamics evolution caused by varying the skewness parameter is hidden in the properties of the L{\'e}vy noise. When parameter $\beta$ possesses positive values, the maximum of the noise PDF shifts to the left: to the area of negative $\xi$ [Fig. \ref{fig8}]. As a result, the most probable values $\xi(t)$ (the values close to to the peak of the PDF in Fig. \ref{fig8}) are negative and the effective parameter value $a_{\text{eff}}=a+\xi_{\text{peak}}$ ($\xi_{\text{peak}}$ corresponds to the peak of the PDF) becomes lover. Changing $a_{\text{eff}}$ has the same effect as varying parameter $a$ and approaching the Andronov-Hopf bifurcation in the deterministic oscillator (see Eqs.(\ref{eq:system}) at $\xi(t)\equiv 0$). It is known that coherence resonance in model (\ref{eq:system}) is more pronounced when values of parameter $a$ obey the condition $|a|>1$ but becomes closer to the critical value $a=\pm1$. In the FitzHugh-Nagumo oscillator subject to the L{\'e}vy noise this effect is achieved due to the noise asymmetry. Similarly, negative values $\beta$ result in suppression of coherence resonance due to increasing $a_{\text{eff}}$ and shifting away from the Andronov-Hopf bifurcation value.

\section{\label{sec:conclusion}Conclusion}
It has been established that the properties of L{\'e}vy noise provide for controlling the noise-induced dynamics of excitable systems. The features of the L{\'e}vy noise influence has no principal impact on the shape of spikes, but are reflected in the evolution of the correlation time and of the interspike interval deviation coefficient observed when the noise parameters are tuned. In particular, the impact of L{\'e}vy noise on the phenomenon of coherence resonance is characterised by two distinguishable aspects. The first one is the influence of L{\'e}vy jumps in the absence of the noise skewness resulting in suppression of coherence resonance and decrease of the noise-induced oscillation regularity (coherence). In such a case, the ability of spiking persists, but the level of the interspike interval deviations increases when the noise signal more and more differs from the Gaussian noise. 

The second revealed peculiarity of the L{\'e}vy noise impact is the possibility to enhance or to suppress coherence resonance by varying the skewness of noise. This effect is associated with the transformations of the noise signal reflected in changing the distribution of the probability density when tuning the skewness parameter. Here, the terms 'enhancement' and 'suppression' are used in the context of comparison with a case of zero skewness of non-Gaussian noise (for the noise stability index less than two). It is important to note, that coherence resonance induced by the L{\'e}vy noise and enhanced by the noise skewness is similarly pronounced as compared to the same effect induced by Gaussian white noise.

The coherence resonance control was demonstrated both in numerical and electronic experiments and the numerical and experimental results are in a good correspondence. The presented results clearly indicate that the L{\'e}vy noise-induced dynamics is more diverse as compared to a classical, Gaussian stochastic force. However, the obtained materials are the first step towards understanding the L{\'e}vy noise-induced coherence resonance and open a list of questions. Has the impact of the L{\'e}vy noise on coherence resonance associated with type-I excitability the same character? Can one control the effect of coherence resonance in non-excitable systems by varying the L{\'e}vy noise parameters in the same way? Does the L{\'e}vy noise allow to control the collective dynamics of coupled excitable elements? These and other problems are the issues for further investigations. 

\section*{Declaration of Competing Interest}
The authors declare that they have no known competing financial interests or personal relationships that could have appeared to influence the work reported in this paper.

\section*{Data Availability}
The data that support the findings of this study are available from the corresponding author upon reasonable request.

\section*{Acknowledgements}
Ivan Korneev and Vladimir Semenov acknowledge support by the Russian Science Foundation (project No.  23-72-10040).

%

\end{document}